\documentclass[]{article}

		\usepackage{booktabs} 
		
		\usepackage{pdflscape}
		
		\usepackage[ruled]{algorithm2e} 

		\usepackage{array,graphicx}
		\usepackage{multirow}

		\SetAlFnt{\small}
		\SetAlCapFnt{\small}
		\SetAlCapNameFnt{\small}
		\SetAlCapHSkip{0pt}
		\IncMargin{-\parindent}

		\usepackage{varwidth}

		\usepackage{array}
		\usepackage{amsmath}
		\usepackage{lscape}

		\newcommand{\ignore}[1]{}

		\usepackage{multicol, blindtext}

		\usepackage[utf8]{inputenc}
		\usepackage[T1]{fontenc}
		
		\usepackage[stable]{footmisc}
		
		\usepackage{enumitem}
	\usepackage{lipsum}
	
\begin{document}
		
		\pagestyle{plain}
		\pagenumbering{arabic}
		

		\author{Katarzyna Kapusta, Gerard Memmi \\ Télécom ParisTech Université Paris-Saclay, LTCI \\ \{katarzyna.kapusta, gerard.memmi\}@telecom-paristech.fr}
		

		\title{Data protection by means of fragmentation in various different distributed storage systems - a survey}

		\begin{abstract}
		This paper analyzes various distributed storage systems that use data fragmentation and dispersal as a way of protection.
		
		Existing solutions have been organized into two categories: bitwise and structurewise. Systems from the bitwise category are operating on unstructured data and in a uniform environment. Those having structured input data with predefined confidentiality level and disposing of a heterogeneous environment in terms of machine trustworthiness were classified as structurewise. Furthermore, we outline high-level requirements and desirable architecture traits of an eficient data fragmentation system, which will address performance (including latency), availability, resilience and scalability.
		\end{abstract}

		\maketitle

		
		\section{Introduction}

		Outsourcing data storage to cloud-based services gains popularity among all kinds of users, as it seems to be a reasonable alternative to a private cloud.
		Cost aside, customers subordinate their choice of an adequate cloud provider to various factors, particularly availability, security, and privacy of the stored data.
		
		Presented analysis focuses on one particular category of distributed storage solutions: systems using data fragmentation and dispersion not only for availability, but also as a way of providing additional data security. Indeed, data fragmentation is widely used for resilience and scalability purposes:  it can be found in the RAID technology \cite{bib:raid}, as well as in clusters like Apache Hadoop \cite{bib:hadoop}. However, the act of dividing data into fragments and dispersing these fragments over multiple machines in order to discourage an attacker is not yet widespread.
		
		Fragmentation for security is not a new idea. It can be found in Adi Shamir's seminal paper from late 70s \cite{bib:shamir}, where he addresses the problem of secure storage and management of an encryption key. A more architectural Fragmentation-Redundancy-Scattering technique was proposed in the 80s \cite{bib:frs1} with a design separating sensitive data from non-confidential fragments and dispersing them on separate physical devices. In the following decades, the idea of fragmentation was applied to relational database systems~\cite{bib:aggarwal,bib:ciriani} and more recently, it was used in the context of cloud or even multi-cloud architectures allowing data separation~\cite{ bib:bkakria, bib:bohli, bib:hudic, bib:vimercati}. Today, we are facing new challenges along scalability and public exposure: petabytes of data are to be protected over large distributed systems of thousands of machines, some stored in public areas for obvious reasons of cost reduction. Information dispersal appears to perfectly fit such multi-tenant environments~\footnote{\tiny http://wikibon.org/wiki/v/3\_Tenants\_of\_Security\_and\_the\_Role\_of\newline\_Information\_Dispersal
		}. \ignore{Moreover, not only data, but also application logic may benefit from partitioning~\cite{bib:bohli}.} 
		
		This survey significantly expands the early work published in~\cite{bib:kapusta} and~\cite{bib:memmi_kapusta_han}  with more precise comparisons and descriptions including more systems and references. Existing storage systems using fragmentation for security purpose are organized into two categories: bitwise and structurewise. The first group  addresses the need for archiving data structures on storage architecture without making any assumptions about the type of data or the kind of storage servers (they are all considered identical in terms of trustworthiness). This group contains not only academic proposals, but also novel commercial solutions. In the second group, confidential data are stored on secured devices in contrast with non-sensitive data that can be kept in public areas. Both categories of systems are analyzed, with regards not only to security but also to memory use, data resilience, key management and performance (including latency).

		{\bf Outline} This paper is organized as follows. Section~\ref{sec:brute-force} contains an analysis of bitwise systems. Section~\ref{sec:structured} describes structurewise solutions. A discussion on some fragmentation issues and recommendation with an insight into possible future research tracks ends the survey.
		
		\section{Bitwise data fragmentation and dispersion}
		\label{sec:brute-force}
		
		This section presents an overview of most notable bitwise fragmentation techniques and systems. Section~\ref{sec:secrecy} describes data secrecy techniques based on fragmentation. Section~\ref{sec:redundancy} to Section~\ref{sec:deduplication} portrays other important characteristics and elements proper to bitwise fragmentation systems: data redundancy, key and fragments' location management, integrity, data defragmentation, trusted area, decoys, fragment size, and data deduplication.
		
		Section~\ref{sec:systems} contains individual descriptions of eight selected bitwise fragmentation storage systems. It treats not only academic, but also commercial solutions. Table~\ref{table:tabelka1} (containing academic solutions: PASIS\cite{bib:pasis}, POTSHARDS\cite{bib:potshards}, GridSharing\cite{bib:subbiah}, DepSky\cite{bib:depsky} and CDStore\cite{bib:cdstore}) and Table~\ref{table:tabelka2} (describing commercial products: IBM Cloud Object Storage (previously Cleversafe{\small{\small\textregistered}}) \cite{bib:resch}, Symform\cite{bib:tabbara1,bib:tabbara2,bib:tabbara3} and SecureParser{\small\textregistered} \cite{bib:unisys1,bib:unisys2,bib:unisys3}) summarize their features with regard to secure data dispersion: secrecy, key management, availability, integrity and recovery in case of loss of information, defragmentation processing, location of the trusted element, number of fragments sufficient for an attacker to recover data, and required storage space. In order to unify systems' descriptions, adequate notation and terminology were introduced in Table~\ref{tab:terminology}.
		
		
		
		
		\begin{table}[h]
		\caption{Notation and Terminology}
		\label{tab:terminology}
		\centering
		\def\arraystretch{1.5}
		\begin{tabular}
		{
		|>{\raggedright\arraybackslash}p{0.15\linewidth}|>{\raggedright\arraybackslash}p{0.75\linewidth}|}
		\hline
		   $d$ & initial data \\ [0.02cm] \hline
		   $d_{size}$ & size of the initial data \\  [0.02cm]  \hline
		   $n$ & total number of fragments corresponding to one initial data \\  [0.02cm]  \hline
		   $k$ & number of fragments required for the recovery of initial data \\  [0.02cm]  \hline
		   $p$ & (ramp schemes) number of fragments not revealing any information about the initial data \\  [0.02cm]  \hline
		   data chunk  & non-processed part of initial data, usually consecutive bytes of data \\ [0.02cm] \hline
		   data share  &  protected piece of data, typically an encoded data chunk \\ [0.02cm] \hline
		   fragment     & final fragment composed of data shares and stored in one physical location \\ [0.02cm] \hline
		   map     &  information about the mapping between data and fragments, fragments' location and, if applicable, the order of fragments \\ [0.02cm] \hline
		\end{tabular} 
		\end{table}
		
		
		\subsection{Data secrecy}
		\label{sec:secrecy}
		
		During bitwise fragmentation processing initial data $d$ are transformed into $n$ unrecognizable fragments that will be stored over $n$ different physical locations. Data reconstruction is only possible when $k$ of these fragments have been gathered. Therefore, resulting data protection depends mainly on the two parameters $k$ and $n$, that define the dispersion scope. A higher value of $k$ implicates that the work required for an attacker to get an access to at least $k$ storage locations will be harder. A value of $n$ close to $k$ increase availability, but makes this work even harder, as the choice of fragments is being limited. In a particular case, when $k$ equals $n$, all fragments are needed for data recovery. Such fragmentation processing is called an {\it all-or-nothing} transform.
		
		Data dispersal is one obstacle on the way to data recovery, but the overall data protection depends also on the protection strength of the algorithm used for transformation of initial data into a set of corresponding fragments. This second parameter corresponds to the difficulty level of obtaining some information about the initial data from less than $k$ fragments. \textit{Information-theoretic (or perfect) security}  provides the highest level of secrecy: no information at all can be deduced from an encrypted data fragment. In the case of \textit{computational security}, an attacker disposing of enough time and computational resources can deduce some information from fewer fragments than the minimum amount required for data reconstruction. When it is possible to obtain some information about the encrypted data we talk about the lowest, \textit{incremental}, level of security.
		
		Three categories of algorithms may be associated with these three protection levels. The first category contains secret sharing techniques providing information-theoretic security. The second one gathers techniques computational secret sharing methods, especially all fragmentation algorithms based on symmetric encryption. Information Dispersal Algorithms (IDA) and straightforward techniques like decimation, characterized by a lower - often insufficient to be used as the only protection - confidentiality level, form the third group. 
		
		Following subsections describes in more details each of the three categories. Table~\ref{table:techniques} gathers the most important in the context of distributed storage fragmentation techniques from all categories and compares their protection strength, performance and capacity of providing data resilience in addition to data security.

		
		\subsubsection{Perfect secret sharing schemes} 
		
		Perfect (or information-theoretically secure) secret sharing schemes transform data $d$ into $n$ fragments, each of a size equal to the size of $d$. Any $k$ of these fragments suffice to recover original information, while $k-1$ fragments do not provide any information about the initial data, even if an attacker possesses unlimited  computational resources. Large fragment size and in consequence huge storage overhead are a serious disadvantage of perfect secret sharing. Therefore, techniques from this category are often judged unpractical to be applied on voluminous data and rather used for protection of small data, typically encryption keys. Nevertheless, three of five academic systems from Table~\ref{table:tabelka1} adopt perfect secret sharing for data protection, judging the increase of storage as an acceptable cost. POTSHARDS and GridSharing chose XOR splitting because of its fast performance. PASIS accepts the use of perfect security in some cases. However, the system does not predefine its data protection technique: once the nature of the data is known, an appropriate algorithm is chosen from a range of perfect, incremental and computational threshold schemes. Details of a mindful choice of a data distribution scheme inside PASIS were described in ~\cite{bib:pasis-scheme}.
		
		
		\textbf{Shamir's secret sharing} A ($k,n$)-threshold Shamir's secret sharing scheme (SSS)~\cite{bib:shamir} takes as input data $d$ and transforms them into $n$ fragments $f_1,...,f_n$, of which at least $k$ are needed for data recovery. The algorithm is based on the fact that given $k$ unequal points $x_1,...,x_k$ and arbitrary values $f_1,...,f_k$ there is at most one polynomial $y(x)$ of degree less or equal to $k-1$ such that $y(x_i)=f_i, i=1,...,k$. In more details, Shamir's scheme uses modular arithmetic. The set of integers modulo a prime number $p$ forms a field in which interpolation is possible. In order to encode data $d$ a prime number $p$ is picked, which is greater than $d$ and $n$. The coefficients $a_1$,..,$a_{k-1}$ of the polynomial $y_d(x)$ are randomly chosen from a uniform distribution over the integers in $[0,p)$ and the values $f_1$,..,$f_n$ are computed modulo $p$.
		
		In SSS, the fragmentation consists of computing a value of the polynomial $y_d(x)$ at $n$ points $x_1,...,x_n$. The complexity of computing a value at single point is $O(n)$. Therefore, it takes $O(n^2)$ to compute a polynomial of degree $n$ at $n$ points. To defragment data we need to recover the constant term of the polynomial $y_d(0)$. This is also an operation quadratic in function of $k$. In case of larger data, Shamir advises to break the initial data into a set of smaller data chunks and apply the fragmentation processing to each chunk separately. Moreover,  implementations of the scheme usually optimize its performance for larger data by performing all of the operations in a finite field $GF(2^8)$, adopted to the nature of byte computations\footnote{\tiny http://manpages.ubuntu.com/manpages/xenial/man7/gfshare.7.html}.
		
		Shamir's scheme was primarily designed to protect encryption keys. In this use case, its quadratic complexity and huge storage overhead are acceptable, but in the context of storage of larger data they may be a serious burden. Nevertheless, the scheme was taken into account as one of the protection method inside the PASIS system. Moreover, POTSHARDS used it as a way to add redundancy to already protected data fragments. 
		
		Around the same time than Shamir, Blakley~\cite{bib:blakley} published his own scheme relying on the fact that any $n$ nonparallel {\it(n-1)}-dimensional hyperplanes intersect at only one specific point. However, it did not find application inside distributed storage systems.
		
		
		\textbf{XOR-splitting} XOR-splitting is an easy and very fast way of implementing an all-or-nothing scheme. The technique relies on the one-time pad encryption technique~\cite{bib:shannon}. In order to obtain $n$  final data fragments, $n-1$ random fragments are generated and one additional fragment is calculated, which XOR-ed with other fragments will give the initial data back. In contrary with other secret sharing schemes found in distributed systems, this method does not provide data redundancy in addition to secrecy. In order to achieve resilience of protected data, XOR-splitting has to be combined with another technique, like data replication (GridSharing), information dispersal or a $(k,n)$-threshold secret sharing (POTSHARDS).
		
		
		\textbf{Ramp schemes} Blakley and Meadows (1985) ~\cite{bib:ramp} first introduce ramp schemes, in which data are broken into $n$ fragments, such that any $k$ of them allows data recovery and fewer than $p$ fragments reveal no information at all. The main idea behind ramp schemes is to gain on efficiency by relaxing on security requirements. The linear ramp scheme, one of the most important type of ramp schemes,  takes as input $m$ data chunks of initial information and $k-m$ other predetermined types of inputs (some of them may be random). From these two inputs it produces $n$ final fragments in a way that the $m$ data chunks can be reconstructed from any $k$ of the fragments (this can be achieved by applying any information dispersal algorithm). In the context of distributed storage, ramp schemes have been considered by the PASIS system. Li, Qin, Lee, and Li (2001)~\cite{bib:convergentdispersal} proposed Convergent Ramp Secret Sharing Scheme (CRSSS), a modification of the linear ramp scheme conceived to be used for data dispersal in a cloud-of-clouds. CRSSS replaces random information inside a classical ramp scheme with deterministic hashes generated from the initial data. Such processing allows further fragments deduplication (deduplication is described in details in Section~\ref{sec:deduplication}).

		
		\subsubsection{Computationally secure algorithms}
		
		Computational secret sharing methods (including all application of symmetric key encryption to fragmentation) provide with a lower protection level than perfect secret sharing, but significantly optimize the size of produced fragments.
		
		\textbf{Secret Sharing Made Short} In \cite{bib:krawczyk}, Krawczyk introduces \textit{Secret Sharing Made Short} (SSMS): a space-efficient way of protecting information using a combination of perfect secret sharing and data dispersal. Initial data are encrypted with a randomly generated key and then transformed into $n$ fragments using an information dispersal algorithm (IDA) (see~\ref{subsec:incremental} for more about IDAs), $k$ of which are needed for data reconstruction. The key is fragmented using a perfect $(k,n)$-threshold scheme and such $n$ key fragments are attached to previously produced $n$ data fragments. To recover data, a user has to gather $k$ fragments. He reconstructs the key using perfect secret sharing, recovers encrypted data from $k$ fragments using information dispersal and then decrypts data with the recovered key. SSMS does not only optimize the storage blow-up, but has a better performance than Shamir's scheme. It is still considered as one of fragmentation methods inside modern cloud-base solutions \cite{bib:future}. However, most systems chose to combine encryption with systematic error correction instead of information dispersal, as it allows an even more faster processing\cite{bib:resch}. For instance, Symform divides (in a straightforward way) encrypted data in 64 fragments and adds then 32 redundant fragments using error correction. A more sophisticated combinations of encryption and resilience make use of \textit{all-or-nothing} transforms. 
		
		\textbf{AONT-RS and CAONT-RS} The \textit{all-or-nothing} (AONT) encryption mode was first proposed by Rivest (1997) \cite{bib:rivest}. It consists of two steps: a package transform followed by an ordinary encryption. The package transform takes as input initial data divided into blocks ( same blocks as during a block cipher encryption) and transforms the blocks in a way that it is not possible to recover them when even one of the transformed block is missing. This is achieved by encrypting the blocks with a pseudo-random key and producing one additional fragment that is the exclusive-or of the pseudo-random key and a hash for each transformed chunk. Resch and Plank (2011) adopted a variant of the AONT inside their AONT-RS technique. AONT-RS first transforms data into $k$ fragments using a all-or-nothing transform and then uses a systematic version of Reed-Solomon error correction coding to produce $n-k$ redundant fragments. The all-or-nothing transformation inside AONT-RS slightly differs from the package transform initially proposed by Rivest. First modification, a known value is appended to data prior to encryption, so it is possible to check data integrity and it is not possible for an attacker to corrupt data while having less than required threshold of fragments. Second difference is that the hash function is applied only once over the whole initial data and not one by one on single blocks of data. This improves the performance of the technique and allows the use of both, block and stream ciphers. Data transformed using AONT-RS are not envisaged to be encrypted a second time. Therefore, a brute-force attack on the random key used for the AONT pass can be envisaged leading to he decryption of the fragments in the possession of an attacker. AONT-RS is applied for distributed storage inside the IBM Cloud Object Storage.
		
	Li et al. (2001) introduced CAONT-RS, a modified version of AONT-RS that allows fragments' deduplication. In CAONT-RS the random key used for the AONT transform is replaced with a cryptographic hash generated from the initial data. A version of CAONT-RS was employed inside the CDStore system. Its AONT processing step uses \textit{optimal asymmetric encryption padding} (OAEP)\cite{bib:oaep,bib:securityofoaep} instead of a block cipher encryption, but the rest of processing remains the same.
	
	SecureParser{\small{\small\textregistered}} also makes of AONT, but instead of applying it to data it uses it to protect the keys. The kind of the AONT is not publicly specified.


	\subsubsection{Incremental security}
	\label{sec:incremental}
	
	The last category of algorithms contains mainly the family of information dispersal algorithms. Other methods, like for instance data decimation (just mentioned by PASIS) did not attract attention in the context of distributed storage.
	
	 \textbf{Information dispersal algorithms} Rabin (1989)~\cite{bib:rabin} was the first to introduce the concept of an information dispersal algorihm (IDA). An IDA fragments data of size $d_{size}$ into $n$ fragments of size $\frac{d_{size}}/k$ each, so that any $k$ fragments suffice for reconstruction. In more details, the dispersal is realized by multiplying initial data (in form of a vector of size $k$) by a $k \times n$ nonsingular {\it generator matrix} that generates $n$ data fragments. Data recovery process is the direct inverse of fragmentation: any $k$ fragments are multiplied by the inverse of the {\it generator matrix}. Rabin's IDA is mainly used for fault-tolerant storage and information transmission. For security applications, it is usually considered as not being sufficient as the only protection. Indeed, a pattern appears when all vectors of data are fragmented using same {\it generator matrix}. A similar problem occurs when using the Electronic Code Book block cipher mode for block cipher encryption~\cite{bib:ecb}. Nevertheless, when applied on already encrypted data, IDA adds redundancy and strengthen protection (like in previously described Krawczyk's scheme). 
	 Li (2012)~\cite{bib:ida-conf} analyzed in a more precise way the confidentiality of IDAs. He considers that Rabin's IDA has {\it strong confidentiality}, because the original data cannot be explicitly reconstructed from fewer than the $m$ required fragments, contrarily to some {\it weak confidentiality} IDAs where it is possible to recover plaintext from less than the required amount of fragments. Furthermore, he presented an effective way of constructing an IDA from an arbitrary $k$ of $n$ erasure code.
	 
	IDAs confidentiality level is quite low, but they have (in some cases) advantages over other techniques. For instance, their space efficient size of fragments make them more suitable than perfect secret sharing for protecting data in motion as it limits the bandwidth usage. Computational secret sharing provides similar fragment size and better confidentiality than IDAs, but for low values of $k$ IDAs show better performance\cite{bib:resch}.
	Recently, several works focused on a way of providing a lightweight and space efficient fragmentation algorithm that would get rid off the pattern preservation problem of incrementally secure algorithms. Cincilla, Boudguiga, Hadji, and Kaiser (2015) \cite{bib:systemx} proposed a mechanism for partitioning data in a multi-cloud environment at a lower time overhead than classical cryptographic techniques. The core of its processing is based on  data shredding combined with a data pattern elimination technique. Kapusta, Memmi, and Noura (2016) introduce a keyless efficient algorithm for data fragmentation that overcome the data pattern problem~\cite{bib:kapusta-ccs}.	
		
	\subsection{Data redundancy}
	\label{sec:redundancy}
		
		A resilient distributed storage system should ensure data redundancy, as it has to be prepared for the lost or alteration of a part of its data in case of an attack or incident. The exact amount of $n-k$ redundant fragments should depend mainly on the trustworthiness of the machines and planned longevity of the system. Indeed, data dispersed over unreliable machines (like in a peer-to-peer system) are more likely to be lost or altered. At the same time, it is easier to plan an efficient data protection strategy if the longevity of the system is measured in years rather than decades.
		
\begin{figure}[ht]
		\centering
		\includegraphics[width=0.80\textwidth]{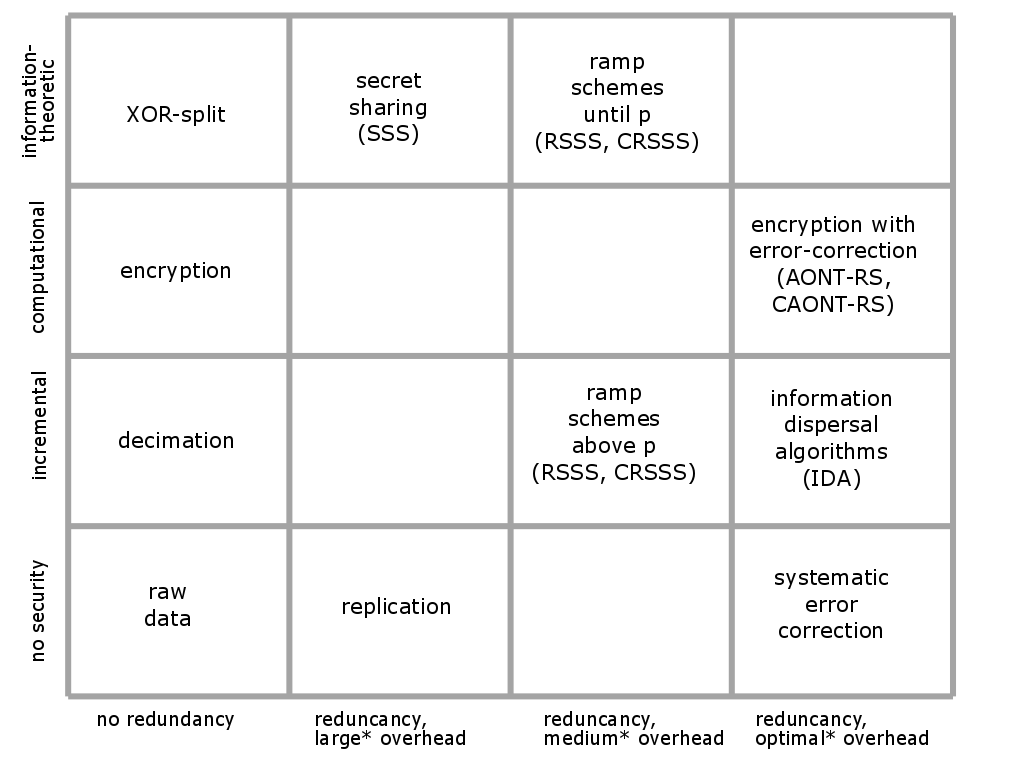}
		\caption{Comparison of redundancy techniques used in {\it bitwise} storages systems based on fragmentation.}
		\label{fig:redundancy}
	\end{figure}

		All of analyzed systems add redundancy to the stored data in order to countermeasure integrity and availability problems. As presented in Figure \ref{fig:redundancy}, the performance of techniques producing data redundancy varies from slow (sharing schemes) to fast (data replication). Nevertheless, the choice of the best technique is not straightforward: not only performance, but also other factors, like the impact on data protection or the increase of the required storage capacity, have to be taken into account. 
		
		Replication is the easiest and fastest solution, but also the most inefficient in terms of memory occupation. It was chosen by GridSharing, where data fragments are replicated over multiple servers and for each of user requests several replicas of the same fragments are retrieved. This strategy protects against data byzantine-faulty storage servers, as the user will choose the replica that appeared in the majority of answers. IBM Object Cloud Storage also applies replication, but only on smaller files, where the gain in performance prevails over the storage blow-up.
		 
		$(k,n)$-threshold schemes and information dispersal algorithms provides not only data privacy, but also redundancy. In the case of perfect secret sharing, strong data protection with additional data resilience comes at a cost of huge increase of memory, same as for replication. Moreover, the performance of such schemes is much slower than of the other techniques. Shamir's secret sharing was adopted by POTSHARDS  long-term storage system. Data fragments split using XOR-splitting techniques are fragmented one more time using the Shamir's scheme. Such combination provides very high data protection at a cost of large memory use. For systems other than archival it may be judged too excessive. Information dispersal algorithms (IDA) solve the problem of storage blow-up, but their security level is rarely sufficient. Therefore, they are applied in addition to another method, like in Krawczyk's scheme where encrypted data are fragmented using an IDA. Such double processing is less efficient than symmetric error correction, where redundant fragments are added to already existing ones. In consequences, except DepSky, all of the systems using symmetric encryption for data protection have chosen error correction (different versions of the Reed-Solomon code) to generate data redundancy.

		\subsection{Performance}
		\label{sec:performance}
			


\begin{figure}[ht]
		\centering
		\includegraphics[width=0.90\textwidth]{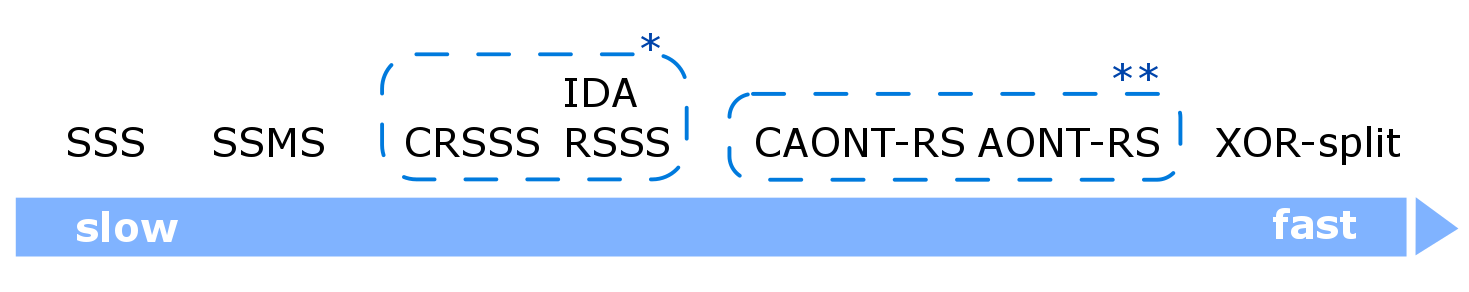}
		\caption{Performance comparison of bitwise fragmentation techniques.}
		\label{fig:arrow-performance}
		\end{figure}

The overall performance of a bitwise fragmentation system depends on various elements. Two more factors have to be taken into consideration in comparison with a classical distributed storage system: the performance overhead of fragmentation and defragmentation processing, as well as the latency delay caused by data dispersal.	
	A rough analysis of performance of bitwise fragmentation techniques is shown in Figure~\ref{fig:arrow-performance} based on benchmarks found in the literature~\cite{bib:resch,bib:future,bib:cdstore,bib:convergentdispersal}. Two information-theoretic algorithms are on the two ends of the comparison: Shamir secret sharing is the slowest algorithm in contrary to the XOR-split that is the fastest method. The rest of the algorithms may be regrouped into two groups: a first one containing techniques using information dispersal (IDA, RSSS, CRSSS) and a second gathering techniques mixing symmetric encryption with systematic error-correction (AONT-RS, CAONT-RS). A comparison of performance and complexity of these two groups is not obvious as it depends on several parameters like the number of fragments $k$, as well as the exact choice of the hash and encryption algorithms inside the AONT transform. In a general manner, for large values of $k$ (approximately above 10) the second group outperforms the first one, while for smaller values of $k$ information dispersal seems to be a more efficient solution. Inside the groups the classification is easier. Convergent versions of algorithms are slower than the original ones, as they require additional cryptographic hash operations. However, the most recent version of CAONT-RS improves the performance of AONT-RS by replacing AONT based on AES with an AONT based on OAEP~\cite{bib:oaep, bib:cdstore}. Krawczyk's SSMS proposal is not precising the details of the techniques that it exploits. Its implementation that uses Rabin's IDA and Shamir's scheme  is faster than secret sharing and slower than information dispersal~\cite{bib:future}.
	Performance burden caused by latency is inevitable: the risk of delays increase with the dispersion scope. Several techniques of dealing with latency issues have been proposed: parallelization of the fragmentation processing, over-requesting (PASIS), keeping track of the most responsive storage nodes (PASIS, IBM Cloud Object Storage). 
[TODO:more about lattency?]
	
		\subsection{Key and fragments' location management}
		\label{sec:management}
		
		In a secure distributed storage system encryption keys are usually stored inside one or more trusted nodes [TODO: add reference]. Inside bitwise fragmentation systems a classical key management store is not always necessary. The fragmentation processing often does not require any key (i.e, secret sharing) or it disperses the key within the fragments (i.e, SSMS, AONT-RS).  However, in such situation the key is somehow replaced by a {\it map} - a piece of information containing the location and order of the fragments and mapping fragments to their corresponding data. Even if it is less critical than a key, this \textit{map} should be stored in a trusted location or distributed over different nodes, as its possession facilitates the work of an attacker. Giving the \textit{map} to the data owner is a risky solution, as the probability that the owner will loss the map increases with the supposed data longevity~\cite{bib:potshards}. 
		
	From all of the described bitwise fragmentation systems only Delta-4 and Symform do not process keys and store them in a trusted part of the system (Cloud Control for Symform and security sites for Delta-4). The rest of the systems use keyless techniques or fragments the keys in addition to data.
	In all the cases some metadata has to be added to the fragments, so the systems can identify the fragments. Then a \textit{map} with references to the fragments is stored in one trusted place, distributed over several nodes of the system, or given to users.  Encrypting fragments' metadata enforces data protection~\cite{bib:delta4-1}. An interesting design is proposed by POSHARDS, where not only the user possess the \textit{map} of his data fragments, but a sort of distributed \textit{map}, called \textit{approximate pointers}, is attached to the fragments. Thus, data recovery is possible even if the \textit{map} was lost. However, \textit{approximate pointers} are constructed in a way that the emergency recovery takes much more time than the standard procedure, as well as require access rights to considerably more storage nodes. This protects against the use of the approximate pointers by an attacker. The procedure of data recovery using approximate pointers is described in more details in Section\ref{ssec:potshards}.
		
		
		

		\subsection{Data integrity and authentication}
		\label{sec:integrity}
		
	 Dispersed fragments may be altered, especially when they are stored on untrusted devices. Ensuring data integrity can be realized in several ways .
	The most common solution is to use cryptographic hash algorithms to add digests to data before the are being fragmented. It is then possible to verify the integrity of defragmented data by computing their hash and comparing it with the one that was received within the fragments. Data integrity verification based on hashes is present in majority of academic systems. Three of them use signatures (POTSHARDS, DepSky) or the keyed-hash message authentication code (SecureParser{\small\textregistered}) to provide not only data integrity, but also authenticate the fragments. Digests are usually dispersed within the fragments, but it is not always the case. In Symform the encryption key required for data defragmentation is generated from the data hash. Keys (and integrity values at the same time) are then stored in a central trusted element, as they have to be protected.
	
	Ensuring the user does not receive corrupted or incomplete fragments can be also realized basing on a voting system (GridSharing). Fragments are replicated over several nodes. During defragmentation process, several replicas of the same fragment are retrieved. From all the replicas that were fetched, the most frequent answer is chosen. Such solution is quite inefficient in terms of storage capacity and lowers the dispersal protection (replicated data are more exposed), but may be faster as no hash computation is required.
		
	Last approach to data integrity requires an auditing of the distributed system (PASIS).  Activity on storage nodes is tracked down and all  recorded changes are verified. In the case of a data loss or undesired data modification, the state of the node will be reverted.
		
		\subsection{Data defragmentation}
		\label{sec:defragmentation}
		
		Defragmentation is the processes of reconstructing initial data. Two scenarios may be distinguished. In the first one, the location of the fragments is known, unless a major failure occurs in the system. Once the fragments are localized, the data defragmentation process - including collecting fragments, integrity verification, decryption and fusion - can be performed.
		
		The loss of the map is a critical situation. A straightforward solution to this problem would be to broadcast a request to all of the storage nodes in order to discover fragments location. However, this will work only if fragments have a piece of information attached to them describing their origin and order.
		POTSHARDS is the only system addressing explicitly the issue of map unavailability by integrating a distributed \textit{map} within the fragments in addition to the one given to a user.
		
		It is important to consider latency while designing the defragmentation procedure, as it is even more important than the fragmentation time.
		Indeed, a user can just upload data and do not wait for the fragmentation to be finished in contrary to a situation when they make a request for data and wait for the data recovery.
		
		
		\subsection{Trustworthiness}
		\label{sec:trusted}
		
		Three principal levels of device trustworthiness may be distinguished inside a distributed storage system: \textit{trusted}, \textit{untrusted} and \textit{curious-but-honest}. Trusted devices are physically secured. Usually they belong to the user or they are parts of the storage system. They are used for processing (fragmentation, defragmentation, naming) of the plaintext, as well as for storage of data maps and encryption keys if they exist. In each of analyzed systems there must be at least one element that is trusted: the one performing the transformation of plaintext into fragments. In most of them it is integrated with the client appliance (PASIS, IBM Cloud Object Storage, GridSharing, SecureParser, DepSky, CDStore, Delta-4). In some it is a separate part of the architecture (IBM Cloud Object Storage, Potshards, Symform, Delta-4). Intuitively, the communication to and from this component must be secured: a man-in-the-middle attack would reveal all fragments to an attacker. Solutions for providing data security at motion usually rely on encrypted communication, like the use of TLS/SSL. Delta-4 evokes a less typical idea of obfuscating the transmission by mixing fragments of different data (and additional decoys if needed) into one flow during transmission from user workstation to storage nodes. Curious-but-honest (or semi-honest) is an expression describing storage nodes or data providers that may try to learn as much as possible about the data they are storing, but they will run the protocols exactly as specified. Untrusted elements can not only analyze possessed data, but also deviate from the protocol.
	
		The question of a secure processing is unavoidable in case of outsourced data and concerns both categories of data: unstructured (analyzed in this section) and with uneven confidentiality levels (described in the section \ref{sec:structured}). A way theoretical way of addressing this issue could be seen in the use of homomorphic encryption \cite{bib:gentry}. However, Fully Homomorphic Encryption is currently impractical and Somewhat Homomorphic Encryption has limited applications areas. Multi-party computation seems a much more promising direction for following years~\cite{bib:future-directions}. Searchable encryption may easy
		
		\subsection{Fragment size}
		\label{sec:size}
		
		The knowledge of fragment size opens the door to a potential side-channel attack. For instance, data fragments obtained with perfect secret sharing have the size of the data and fragments coming from computational or incremental techniques have the size of data divided by $k$. This fact may not only indicate to an attacker the data size, but also it can help finding out what fragments are belonging to the same fragmentation result (as fragments of the same size most probably will be part of the same fragmentation result). A way to countermeasure this problem is to define one fragment size inside the system: data are then first divided into chunks that of such size. If an exact division is not possible, data are padded to be a multiple of the given size. Such padding may be seen as a sort of decoy. Choice of the size of a data chunk has to be mindful: a smaller data chunk limits the overhead of padding (especially when a lot of files are smaller than required chunk size), but increase the costs of processing (naming and mapping of fragments). One could also imagine a different type of data division into chunks of various size that would also obfuscate information about the size of stored file.
Beside security reasons, data division into chunks may speed up the system performance as it enables parallelization of processing.The size of fragments in described systems is usually left unspecified. Only Symform defines one exact size of its data fragments (described in \ref{sub:symform}). 

Data chunks are not always constructed from consecutive bytes of the initial data. Delta-4 builds them by decimating the data: one byte of $k$ goes to a one fragment. SecureParser{\small\textregistered} allows the use of a more complex processing in addition to encryption, involving splitting of data at a bit level. A similar bit scrambling transformation was proposed in LightBlind \cite{bib:systemx} as an alternative to data encryption: bits of data are dispersed to different providers in a way that single fragments do not conserve patterns of the original data. 
				
		
		\subsection{Decoys}
		\label{sec:decoys}

		
    \textit{Decoys} are fragments that have the role of misleading a potential intruder. They may be implemented in two ways. First way consist in generating fragments that do not belong to any initial data and mixing them with valid fragments. Such additional portion of data obfuscates the exact amount and size of stored fragments, but increase the storage requirements. An alternative and more efficient way of processing consists in injecting invalid entries to the \textit{map} containing misleading locations of the fragments. Therefore, already existing fragments (but matching different data files) are used as decoys. Such technique was adopted by POTSHARDS, where in the secondary \textit{map} only one entry out of four leads to a valid fragment. A technique employing decoys has been already proposed by the authors of Mnemosyne: a steganographic peer-to-peer storage system providing not only data privacy, but also plausible deniability of its stored data \cite{bib:mnemosyne}.

		\subsection{Data deduplication}
		\label{sec:deduplication}
		
		Data \textit{deduplication} is a technique exploiting content similarity of data to reduce the overall storage space inside a system~\cite{bib:venti,bib:convergentdispersal}. Duplicate copies of same data are stored only once. This may be easily done by identifying data using their hash. The method works especially well for archival systems storing a big amount of similar data, for instance backup files (as consecutive versions of backup files usually do not significantly differ).
		Deduplication in a bitwise fragmentation system prohibits any randomness inside the applied fragmentation technique. Indeed, secret sharing or encryption produce different outputs when applied on the same data, as their fragmentation output depends on a random input element (i.e., random coefficients or random encryption key). In an unmodified form, only information dispersal algorithms allows data deduplication when the dispersal matrix remains the same for all fragmented data. It is however possible to easily adapt a fragmentation technique to deduplication purposes by replacing the random element by deterministic data constructed from hashes of the data. Thus, CAONT-RS and CRSSS techniques were proposed~\cite{bib:convergentdispersal} that modify the AONT-RS and RSSS algorithms.
		The use of deduplication is space efficient, but - when the systems is designed in a naive way - it opens the door to side-channel attacks.
	An attacker eavesdropping the communication between a user and the system nodes may for instance deduce if user's data exists already inside the system. A two stages deduplication, proposed by CDStore, provides a solution to this problem.
		

		\subsection{Systems characteristics}
		\label{sec:systems}
		
		\subsubsection{Delta-4} \label{sss:delta4}
	
	Delta-4~\cite{bib:delta4} was one of the first project to address the need for a dependable distributed storage system, that will resist accidental faults and intentional intrusions. The system environment is comprised of three separated areas: a user, a security and an archive site. The user site is a trusted area composed of users' workstations, where data fragmentation and defragmentation occur during user sessions. The archive site stores dispersed data fragments. The totality of the archival nodes is considered as trusted, but single nodes are individually untrusted: they may be subject to accidental faults and malicious intrusions. Security sites have to be trusted. They handle authentication and authorization of users, as well as store and manage fragmentation keys.
	
	Data processing is based on the \textit{fragmentation-redundancy-scattering} (FRS) technique~\cite{bib:frs1,bib:intrusion-fine-grain}. In a pre-processing step data are cut into data chunks of equal size, that are referred to as \textit{pages}. Padding is applied when the data size is not a multiple of the size of a page. Pages are then encrypted with a chained cipher, in a way that the preceding cipher text is necessary for the recovery of the following plaintext. Data from a ciphered page are then decimated byte by byte over $k$ fragments. Such processing creates dependency between ciphered data inside fragments, so a partial defragmentation is not possible without guessing the missing part of the data. Data integrity is ensured by the use of cryptographic checksums that are added at the beginning of each page. $k$ fragments corresponding to one page are then replicated $n$ times and broadcasted to the archive sites. A pseudo-random algorithm is used to decide if a site will store the given fragment or not by taking into account the relative space at each site. Apart of balancing the available space among different sites, the algorithm obfuscates the actual locations of the replicas. In Delta-4, the replication was chosen because of performance reasons, as thirty year ago CPU cycles were to scarce to efficiently execute error-correction or information dispersal. Since all fragments are sent to all storage nodes reducing the bandwidth by decreasing the size of transmitted data is paramount importance~\cite{bib:pasis-scheme}. An interesting concept of decoy use is evoked: during communication between a user site and the storage nodes, the fragments of a single file are supposed to be drown into a flow of different fragments (if necessary this flow may be artificially created).
	
	Fragments naming is realized at the user sites. A unique name is added to each of the fragments, derived from the fragmentation key, the name of the data file, the index of the page and the index of the fragment. Thus, even if an attacker obtains all the required fragments for a given data file, they will have to find out the right ordering of the data fragments.  The fragmentation key used for naming and data fragmentation is stored using secret sharing by the security site. At the beginning of a user session it is transmitted to the user site. It is not possible for an attacker to find any relevant data inside a user workstation except during a running session.

		\subsubsection{PASIS} \label{sss:pasis}
			
		 The objective of PASIS~\cite{bib:pasis, bib:pasis2} project is similar to the one of Delta-4: designing a storage system capable to handle storage nodes failures and activity of malicious users. System architecture includes PASIS storage nodes (vulnerable to attacks and intrusions) equipped with repair agents and client systems with installed PASIS agents (trusted areas). Storage nodes are self-securing: their repair agents internally version all data and requests for a certain amount of time~\cite{bib:pasis-selfsecuring}. Keeping history information allows the detection of intrusions and prevents intruders from destroying or undetectably tampering with stored data. Systems administrators have a window of time to recognize malicious activity and rebuild the old state of the system using the history pool. A major difference between a self-securing storage like PASIS and a conventional backup system is that the security perimeter in the first one is situated around the storage device.
		 	 
Because of the belief that no single data distribution scheme fits all data types and systems, the fragmentation method in PASIS is not predefined, but adapted to the particular kind of data to be stored and to the current system performance. Details of a mindful selection of a fragmentation scheme for a survivable storage system were described by Wylie et al. (2001)~\cite{bib:pasis-scheme}. Basically, the final choice of a \textit{p-m-n} threshold scheme for a given data file is a trade-off between desired performance, data availability and security. A wide range of techniques and their combinations is taken into account: from replication to error-correction for data availability and from information dispersal to secret sharing for data privacy.	 

A fragment is identified by the name of the storage node and its local name on that node. A dedicated directory service maps the name of data stored over storage nodes to their fragments.  Therefore, a careful naming of stored files can obfuscate relations between fragments. 

PASIS comes up with two suggestions that aim at improving the performance of retrieval of the distributed data. First consists in over-requesting of data fragments: asking storage nodes for more than $k$ fragments during the data retrieval. This way, only the $k$ first fragments that arrived first are defragmented. However, at the same time the bandwidth usage is increased. Second suggestion is to send the requests to the storage nodes that have responded the most quickly during recent retrievals.

		\subsubsection{GridSharing}
		
		GridSharing \cite{bib:subbiah} is another implementation of a distributed storage system, dating from 2005. It combines a perfect sharing scheme with the replication of the secret fragments on different devices in order to build a fault-tolerant and secure distributed system, fulfilling confidentiality, integrity and availability requirements.
		According to the authors, choosing perfect sharing schemes for data protection alleviates problems related to key management.
		
		An interesting idea of possible fragments renewal for additional protection is developed. It would be based on a periodical replacement of existing data fragments with new ones. This way an attacker has a much shorter time to collect all of the fragments required for data recovery. Perfect sharing schemes (where all of the fragments are required to reconstruct the data, like the XOR splitting method) allow for the implementation of efficient share renewal procedures: changing stored fragments without prior data defragmentation.  For imperfect sharing schemes, where possession of fewer fragments than the required threshold can provide some information, a renewal algorithm has not yet been developed.
		
		GridSharing accounts for several ways that the storage servers could fail. They could leak information, revealing their content and state to an adversary, while executing the specified protocols faithfully. They could be Byzantine faulty, deviating from the specified protocol or revealing their fragments. They could also just crash. It introduces a new \textit{l-b-c} scheme, in which up to \textit{l} servers are leakage-only faulty, up to \textit{b} are Byzantine faulty and up to \textit{c} servers can crash.
		\newline \indent
		Subbiah \cite{bib:subbiah} shows high computation overhead of Shamir's secret sharing scheme. In GridSharing, a combination of two mechanisms is proposed in order to overcome inevitable performance problems. First, a XOR perfect sharing scheme is used, where all fragments are needed for secret recovery. Second, fragments are replicated on servers, using a voting system to determine incoherent fragments. For each share, at least \textit{(2b+1)} responses must be received. The value returned by at least \textit{(b+1)} servers is the correct one.
		\newline \indent
		Two fragment allocation schemes for dispersing data over \textit{n} servers are presented. In the first one, called direct approach, servers are arranged in a logical grid of \textit{(l+b+1)} rows, with at least \textit{(3b+c+1)} servers in each row. The data is split in as many fragments as there are rows. Each fragment is then replicated along a single row. In the second approach, named GridSharing, each of the servers contains a couple of fragments. \textit{n} servers are arranged in the form of a logical rectangular grid with \textit{r} rows and \textit{n/r} columns. As in the first approach, servers in the same row replicate the same data.
		\newline \indent
		The GridSharing system comes with stronger security than encryption-based techniques and provides an easy way of sharing data in a collaborative environment. The dimensions of the architectural framework may be altered to trade-off between number of required servers, storage use, as well as recovery computation time.
		
		\subsubsection{POTSHARDS} \label{ssec:potshards}
		
		In 2007, the POTSHARDS \cite{bib:potshards} project addressed the need for providing a secure archive with the ability to last for years. Its basic concept is to distribute data between multiple cooperating organizations forming an archive system.
		
		The authors decided to use secret splitting schemes instead of encryption for two reasons. First, key management can be expensive over years, as it requires key replacements, and there is no guarantee that an encryption key will not be lost. Second, even the strongest encryption is only computationally secure and can become easily breakable over a finite period of time with the development of new technologies (as the objective is to protect data for decades).
		
		POTSHARDS encryption works in a two step process. First, random fragments are generated from user data using XOR-splitting, providing information-theoretical security. Each of those is used to generate a group of {\it shards} -- data pieces of the same length as the initial fragment -- using Shamir's threshold scheme, providing data availability. Shards are then distributed across independent organizations. POTSHARDS assures data integrity by the use of algebraic signatures.
		
		Objects, fragments, and shards can be identified by their IDs. After data distribution, a user obtains a list of indexes corresponding to their archived objects. Because it is possible for this list to be lost, shards include additional portions of information called \textit{approximate pointers}. Pointers of one shard show the archive region, where shards from the same object are located. As consequence, a user can recover data from the shards even if all the other information, such as the index, is lost. An intruder would have to steal all of the shards that approximate pointers refer to. This implies, among other things, bypassing the authentication mechanisms of each archive.
		
		In the event of a partial data loss, the archives can collaborate to recover the missing parts without revealing any information about the encoded content. They start by agreeing on the destination of the data to be recovered, by choosing a new fail-over archive. The missing data is then sent to this archive in several rounds. In a round, each collaborating archive generates a random block and XORs it with a block of data needed for reconstruction. All those encoded blocks are XOR-ed together and the result is sent to the fail-over archive, which also recieves the encoding keys used by the other archives. With all these pieces, it is able to recompute the lost data.
		
		POTSHARDS provides information-theoretical security and does not require any key management. With enough time, it is possible to recover data even if shards location maps have been lost. Collaboration between organizations allows to rebuild a lost archive, but also implies a trade-off between secrecy and reliability. POTSHARDS pays for its goal of providing a long term archive by requiring a large amount of storage: both the secrecy and the availability splits are space consuming.

		\subsubsection{DepSky}

		DepSky~\cite{bib:depsky} is an academical dispersed storage system built on top of four commercial clouds. This cloud-of-clouds aims to improve data availability, confidentiality and integrity. Indeed, dispersal over multiple providers protects against a situation when data become lost or unavailable when the storage provider has been attacked. It is also a Byzantine fault-tolerant, as it is possible to verify if the data have been corrupted and retrieve only the valid data fragments. Moreover, as each of data hosts stores only a portion of data, the system facilitates moving data from one provider to another avoiding the vendor lock-in problem.
		
		The system architecture is composed of a set of clients that read and write data stored over four commercial clouds. To deal with the heterogeneity of cloud interfaces, DepSky data are encapsulated inside special {\it data units}, which exact implementation depends on the architecture of the storage provider. Each data unit contains metadata with a set of information, like data version or signature.
		
		Two protocols for data distribution of data units are proposed: DepSky-A and DepSky-CA. First one does not provide any data confidentiality, but just disperse replicas of data over clouds in order to increase availability. Moreover, as data are replicated the overall storage blowup is equal to number of the replicas. DepSky-CA is a secure and space efficient improvement of the DepSky-A. Data processing follows Krawczyk's SSMS method. First, data are encrypted using symmetric encryption (AES). Encrypted data are then encoded into $n$ fragments by the use of an optimal erasure code (Reed-Solomon). The encryption key is partitioned into $n$ shares using secret sharing (Shoenmakers' Publicly Verifiable Secret Sharing scheme~\cite{Schoenmakers:1999}) and such key shares are attached to data fragments. Data integrity is ensured by the use of $n$ digests, one for each cloud, inside of the metadata (SHA-1 was used for cryptographic hashes and RSA for signatures). The system allows the replacement of the secret sharing by a key distribution infrastructure.
		 
		An analysis of DepSky system demonstrates an improvement of the perceived availability and (in most cases) the access latency, when compared with cloud providers individually. The cost of such data dispersal was estimated to twice the cost of using a single data storage provider.

		\subsubsection{CDStore}
		
		CDStore~\cite{bib:cdstore} addresses the problem of the storage of backup-data in a multi-cloud environment. As versions of backup-data may be very similar between each other, a possibility of data {\it deduplication} (a technique where data with same content are stored only once) optimizes the storage blow-up inside such an archival system. Once initial data are fragmented the CDStore systems stores only fragments that are different from those already archived. Therefore, fragmentation of two identical data has to result into two identical sets of fragments~\cite{bib:deduplication}. To achieve this, a special fragmentation technique named {\it convergent dispersal} ~\cite{bib:convergentdispersal} is introduced that combines keyless security of dispersal algorithm and deduplication. 
		
		The system architecture is composed of CDStore clients on the user side and of one CDStore server per each participating cloud. Initial data are first divided into $chunks$ of variable size. $Chunks$ are then transformed into $n$ fragments using the Convergent AONT-RS technique (CAONT-RS), $k$ of which will be needed for data recovery. CAONT-RS is similar to the AONT-RS technique, except that for performance reasons CAONT-RS replace the Rivest's AONT with a all-or-nothing transformation based on {\it optimal asymmetric encryption padding} (OAEP)~\cite{bib:oaep,bib:securityofoaep}.  Moreover, to allow deduplication instead of a random key, a key created from the hash of the data is used.
		
		The proposed two stages deduplication method resists against side-channel attacks (where an attacker can deduce information about data by seeing only differing fragments being updated to the cloud). First, produced fragments are deduplicated at the client side between themselves. In a second step, the remaining fragments are transfered to the CDStore servers that perform the second deduplication: they will keep only the fragments that are different from those being already inside the clouds. 
		
		Metadata describing initial data is kept inside the CDStore client. Metadata with information about data fragments is stored inside the CDStore server or dispersed over multiple CDStore servers using secret sharing (for a higher level of data security).
		
		\subsubsection{IBM Cloud Object Storage}
		
		IBM Cloud Object Storage\footnote{\tiny https://www.ibm.com/cloud-computing/products/storage/object-storage/} (previously Cleversafe{\small\textregistered}\ \footnote{\tiny http://www.pcworld.com/article/3130792/ibms-cleversafe-storage-platform-is-becoming-a-cloud-service.html}) is one of the first commercial solutions implementing data dispersal as a way to provide security. Its dispersed storage network offers a complete software and hardware architecture for private cloud storage. Petabyte scalability and reliability are the key drivers of the product.
		
		The data is encoded at the source or on dedicated hardware. Users choose the level of redundancy that they wish the fragmentation to produce. Then, the data is transformed and encrypted using the AONT-RS approach \cite{bib:resch}, which combines the All-or-nothing transform \cite{bib:rivest} with Reed-Solomon erasure codes. First, the data is fragmented into same length words, and each word is encrypted with a random key using the AES-256 algorithm. Subsequently, a SHA-256 hash value of the data is generated in order to provide an integrity check. The last word is created by XOR-ing the canary with the key. It is not possible to reconstruct the data unless someone obtains all the fragments and retrieves the key from the last word. For availability purposes, IBM Cloud Object Storage applies a modified version of Rabin's IDA to the encrypted data, based on Reed-Solomon erasure codes. As a consequence, additional pieces of data are generated, so the user can reconstruct the data even if some of the pieces are lost. Fragmented data is dispersed on random storage nodes.
		
		Data must be secured not only in storage, but also during transmission: if an attacker intercepts the traffic and catches all the fragments, he may reconstruct the data. To read the objective of secure transport of fragments, IBM Cloud Object Storage verifies all nodes that would like to join its storage network. Furthermore, data transit is protected by the use of encryption. The lack of need for key management (the key being included in the data) makes data management less costly. Once an attacker gains access to the storage (for example by breaking authentication mechanisms) or succeeds in observing fragments passing through the network, they will have all the elements needed for data reconstruction. With IBM Cloud Object Storage, data protection relies on the inability of an attacker to collect the data from multiple locations or to intercept the traffic of fragments from the storage servers to the client.
		
		A major drawback of the AONT-RS approach is its performance when working with small objects. IBM Cloud Object Storage introduces a separate way of processing such data, which does not use time-consuming error correction codes. Instead, availability of small encrypted data is ensured by replication.
		
		\subsubsection{Symform}
		\label{sub:symform}
		
		Symform\footnote{\tiny http://www.symform.com} uses a peer-to-peer solution to decrease storage costs and to provide additional security measures based on dispersion in addition to standard data encryption.
		
		Symform uses the RAID-96™ patented \cite{bib:tabbara1,bib:tabbara2, bib:tabbara3} technology for data protection and availability. Before being stored in the Symform cloud, data from a dedicated folder on a user device is divided into 64MB blocks and encrypted using the AES-256 algorithm. A unique encryption key for each block is generated from the hash of the block itself. The data is encrypted at the folder level, so this technique allows a de-duplication of data without decryption: if a block already exists in the folder, it is not uploaded again. Furthermore, each block is shredded into 64 fragments of 1MB each. Then, 32 parity fragments are added to every block using Reed-Solomon codes. This results in 96 fragments corresponding to one block of the original data. These fragments are randomly distributed across 96 randomly chosen devices. To reconstruct the protected data, any 64 out of 96 fragments need to be assembled.
		
		The combination of encryption and fragmentation provides strong computational security. In order to decrypt a single block of data, a malicious user would have to locate 64 pieces of information, collect them from the storage devices and then break the AES-256 encryption, for which the key is twice as long as the AES-128 recommended by the NIST. This seems to be an insurmountable effort.
		
		The weak point of the Symform system may be its Cloud Control critical element responsible for management of fragment locations and encryption keys. An attack on the Cloud Control may result in data unavailability.
		
		\subsubsection{SecureParser{\small\textregistered}}
		
		In 2005, Unisys introduced SecureParser{\small\textregistered} \cite{bib:unisys1, bib:unisys2}. This software solution fulfills three requirements of secure data storage: security, integrity and safety.
		It uses physical separation of data and fault tolerance in order to achieve a better security level than traditional encryption techniques.
		
		Data security requirement in SecureParser{\small\textregistered}\ is solved in two ways. The first protection and the core security component is the AES encryption algorithm. Moreover, the data is split at the bit level using a random value that serves as the splitting key.
		This unique processing results in the creation of data shards containing random bits of the encrypted data, which can then be distributed over storage nodes. To avoid key managements problems, both keys (encryption and splitting) are stored inside the data itself using a sort of All-Or-Nothing-Transform.
		
		For better availability, \textit{m} of \textit{n} redundancy can be added to the data before shards distribution. Therefore, the original data can be restored from \textit{m} of all \textit{n} shards.
		Moreover, SecureParser{\small\textregistered}\ introduces the concept of a mandatory share: mandatory shares are required for the proper recovery of the data, regardless of the \textit{m} of \textit{n} specification. It can be interesting in a situation when the user would like to quickly change the fragmented data -
		they will only have to change the mandatory shares. However, it can be seen as a weak point in a situation when an attacker would like to make the data unavailable: to achieve this, they only have to destroy one of the mandatory shares.
		Integrity in SecureParser{\small\textregistered}\ is assured by adding an authentication value to each of shares.
		
		SecureParser{\small\textregistered}\ claims to provice performance gains over other security products, by performing computations on small blocks of data, and by using the AES-CTR mode \cite{bib:aes-ctr}, which allow parallel processing.
		
		\subsection{Academic approach vs. commercial products}
		
		Two distinct trends divide the analyzed systems. The first one aims at optimizing the storage space and performance even at the cost of lower security or more expensive computations. It is adopted by the commercial solutions, which tend to use error correction rather than data replication and AES encryption instead of secret splitting.
		On the other hand, academic systems focus on providing strong, information-theoretic security, combined with high data resilience. Based on secret splitting techniques, they take significantly more storage space. From that trend come novel solutions, such as repair agents in PASIS, or approximate pointers in POTSHARDS, which deserve more experimenting and probing.
		
		Bitwise systems are mainly dedicated to archiving or long-term storage. They aim at maximizing data protection and providing a good data redundancy at the same time. To achieve this goal they combine various techniques like encryption, secret splitting, data replication, error correction and information dispersal. A usual data processing strategy includes two steps: one for security and one for redundancy. The resulting architecture of a storage system is often a compromise between desirable fragment confidentiality level and memory use. Performance and resilience are two other factors that may impact the design of an appropriate data fragmentation algorithm.
		
		\section{Exploiting data structures, multi-level confidentiality, and machine trustworthiness}
		\label{sec:structured}
		
		In case of known data structure, the fragmentation process can proceed by taking into account the varying need for secrecy along different types of data. This way, confidential data can be easily separated from non-sensitive information and storage places of both parts can be appropriately chosen. There is no need to provide a specific secure architecture to store a piece of data that does not reveal anything confidential. This idea was adopted by the authors of the object-oriented Fragmentation-Redundancy-Scattering (FRS) \cite{bib:frs1} technique at the end of the last century.
		Later, it has been modified to suit database storage \cite{bib:aggarwal,bib:ciriani} and cloud computing technology \cite{ bib:bkakria, bib:hudic, bib:vimercati}.
		
		The need of user interaction during the decomposition process is one of the biggest problems of fragmentation of structured data. It is the user's responsibility to provide for each set of data types the rules that define what is confidential and what is not. Still, it remains possible that a combination of two or more non-confidential data fragments will reveal information about the confidential one. This was especially underlined by Ciriani \cite{bib:ciriani}.
		
		A way of separating data in two fragments (confidential and non-confidential) without user interaction was developed in~\cite{bib:han}.

		\subsection{Object-Oriented Fragmentation-Redundancy-Scattering}
		
		In the 90's, fragmentation-redundancy-scattering technique (FRS)\cite{bib:frs1,bib:frs2} was introduce to provide accidental and intentional fault tolerance inside distributed systems with a limited trusted area. Data are divided into non-confidential fragments that are then replicated for resilience reasons and scattered over a large number of nodes. To ensure data protection, the suggestion is to encrypt confidential data before fragmentation and scattering processing. A version of the FRS technique uses threshold fragmentation schemes, like Shamir's, during the redundancy step.
		
		Processing of confidential data has to be done only on trusted sites (on plaintext) or using methods like homomorphic encryption (while operating on the ciphertext). Both ways may be quite costly. Minimizing the amount of confidential data by applying fragmentation at design time would limit the needs for trusted servers or processing power\cite{bib:confidential-objects}. Indeed, while designing an application, confidential objects may be iteratively substituted by a collection of non-confidential objects. This concept was the base for the object-oriented version of the FRS technique\cite{bib:frs-object}.
		In a first step, confidential objects belonging to an application are fragmented using a redundant algorithm until being broken into pieces which do not reveal anything sensitive. Redundancy is achieved by the use of error processing techniques (like error correcting codes) or by anticipating the application design at the early stage of designing objects. At the end, the fragmented data is scattered over various workstations. Leftover fragments, which are still holding confidential information after the first step, are encrypted or stored on trusted devices. All remaining pieces are distributed over untrusted sites. Data processing or defragmentation are performed on trusted sites.

		After two decades, the FRS technique was implemented inside two distributed systems: one peer-to-peer\cite{bib:efrs-kent} and another one with a central server\cite{bib:efrs}.

		\subsection{Database fragmentation}

		Database as a service\footnote{\tiny https://www.technologyreview.com/s/414090/designing-for-the-cloud/} delivers similar functionality to classic, relational, or NoSQL database management systems while providing flexibility and scalability of a hosted in a cloud on-demand platform. The DBaaS user does not have to be concerned with database provisioning issues, as it is the cloud provider's responsibility to maintain, backup, upgrade and handle the physical failures of the database system. It is easy to see that simplicity and cost effectiveness are the biggest advantages of such solution.
		
		However, data owners lose control over outsourced data. This creates new security and privacy risks, especially when databases contain sensitive data, such as health records or financial information. As a consequence, securing database services has become a need of paramount importance. A solution to the problem involves encryption of the data before sending it to the storage provider. Large overhead and query processing limitation are the main drawbacks of such a blunt approach. 
		
		\textit{K-anonymization} \cite{bib:sweeney}, \textit{t-closeness} \cite{bib:li} and \textit{l-diversity} \cite{bib:machana} anonymization techniques can be seen as a special case of multilevel confidentiality. Some progress on this subject, mainly for health data, has been recently performed \cite{bib:bergeat}.
		
\begin{figure}[!t]
		\centering
		\includegraphics[width=0.5\textwidth]{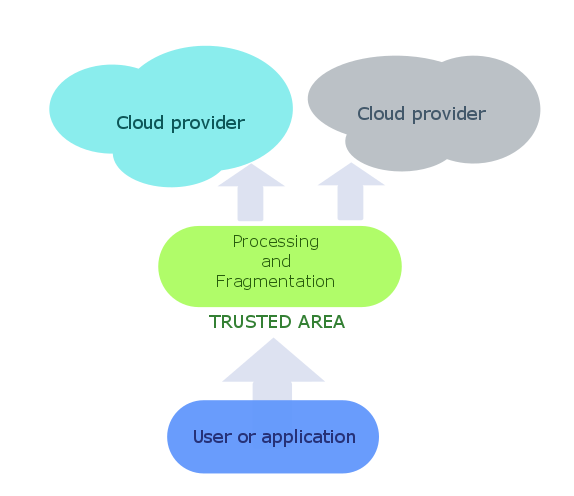}
		\caption{Defragmentation in a trusted zone of information received from two providers}
		\label{fig:trusted}
		\end{figure}

		Database fragmentation promises an interesting alternative to database full encryption or full anonymization. One of the first works on the subject \cite{bib:aggarwal} introduces a distributed architecture for preserving data privacy, while authorizing some processing. A trusted client communicates with the end-users and utilizes two non-trusted servers belonging to different storage providers (see Figure~\ref{fig:trusted}), ensuring the physical separation of the information to be protected. By construction, storage providers do have access to the information that users entrust them with. Even if they are well aware that they should not incorrectly interact with the user’s data and its integrity without endangering their own business, it is a common assumption to suppose them to be \textit{honest but curious}: they have the ability to observe, move, and replicate stored data, especially behind the virtualization mechanism. In \cite{bib:aggarwal}, the outsourced data is partitioned among the two untrusted servers in a way that content at any one server does not breach data privacy. In order to obtain valuable information, an adversary must gain access to both databases. By analogy, the system is also protected from insider attacks and the curiosity of the providers as long as they do not ally together. On top of that, queries involving only one of the fragments are executed much more efficiently than on encrypted data. Nevertheless, this privacy-preserving outsourcing solution has a serious limitation. It assumes that the two servers are not allowed to communicate with each other. In a real life scenario such a condition can be hard to guarantee. The trusted zone is an essential element in Aggarwal's solution. All of the presented systems must include at least one trusted element in their architecture: the one in which the fragmentation and defragmentation processes will occur.
		
		Another work \cite{bib:ciriani, bib:vimercati} protects sensitive information by mixing encryption and fragmentation. It defines confidentiality constraints as a subset containing one or more relation attributes. A constraint involving only one attribute implies that the value of the attribute is sensitive and the only way of protecting it is the use of encryption. On the other hand, multi-attributes constraints specify that only associations between attributes of a given constraint are sensitive. In that case, there is no need to encrypt all the attributes values, because confidentiality can be ensured by fragmentation.
		
		In \cite{bib:ciriani, bib:vimercati}, three scenarios of fragmenting a relation are presented. In the first one, a relation is divided into two fragments, which does not contain sensitive combination of unencrypted attributes. In the second scenario, the relation is split into multiple fragments in a way that any query can always be evaluated on one of the fragments: each fragment contains unencrypted attributes that do not violate confidentiality constraints, as well as the encrypted representation of all other attributes. The last fragmentation scenario avoids the use of encryption by introducing a trusted area (belonging to the data owner) for the storage of sensitive portion of data.
		
		For each scenario, the authors present fragmentation metrics supporting the definition of an appropriate fragmentation algorithm. Fragmentation metrics can aim at minimizing the number of fragments, maximizing affinity between attributes stored in one fragment or minimizing querying costs.
		
		Recently, Bkakria \cite{bib:bkakria} generalized this approach to a database containing multiple relations. It introduces a new confidentiality constraint for the protection of relationships between two tables. Sensitive associations between relations are secured by the protection of primary key/foreign key relationships and the separation of the involved relations. Relations are transformed into secure fragments in which subsets of attributes respecting confidentiality constraints are stored in plaintext, while all others are encrypted. It introduces a parameter for evaluating the query execution cost and proposes a query transformation and optimization model for executing queries on distributed fragments. It also focuses on the issue of preserving data unlinkability while executing queries on multiple fragments. Indeed, providers have to build a coalition and then deduct information by observing query execution. To avoid such situation, \cite{bib:bkakria} proposes the use of an improved Private Information Retrieval (PIR) \cite{bib:olumofin} technique, which allows querying a database without revealing query results to service providers. Results of implementation of the proposed approach are presented. Although the modified PIR solution is much faster than its predecessor, the processing time of record retrieval from multiple fragments is considerably slower in comparison with querying a single fragment.
		
		The idea of splitting a database into fragments stored at different cloud providers was also proposed by Hudic \cite{bib:hudic}. In this approach, a database is first normalized and then several security levels (high, medium, low) are attributed to relations. Based on these three levels and specific user requirements, data is encrypted, stored at local domain or distributed between providers.
		
		Database fragmentation methods presented in this paper remain limited in terms of number of fragments that does not exceed dozens. Moreover, in each case, the proposed fragmentation algorithms require user interaction in order to define data confidentiality level.

		\section{Conclusion}
		\label{sec:conclusion}
		
		In previous sections, we analyzed existing distributed storage systems providing additional secrecy by use of fragmentation. We also presented database fragmentation solutions separating data in order to avoid full encryption. Few systems focus on providing a long term, secure and non-costly data storage. Another motivation is the possibility of minimizing encryption inside databases, while still providing a good level of data protection.
		
		\subsection{Fragmentation: issues and recommendations}
		
		In order to design an efficient storage systems for fragmented data, some problems still have to be overcome.
		
		First, a process of fragment dispersion requires data being separated securely. A situation where data is fragmented, but where we would not or could not control where fragments are stored has to be considered a weak solution. Using multiple, independent providers can be a rapid and coarse-grained solution \cite{bib:hudic}, since it entails significant latency costs.
		
		An alternative could be the storing of the data at a single provider site that guarantees physical separation. Unfortunately, the majority of cloud providers use virtualization, which prevents the end user from such control. With the development of bare-metal \cite{bib:bare} clouds like TransLattice Storm \cite{bib:storm}, Internap \cite{bib:internap} or Rackspace OnMetal \cite{bib:rackspace}, we believe it could be possible to control physical location of outsourced data within a single provider's site.
		
		Second, there are not enough published studies on performance of dispersed storage systems to allow for a comparison with the more common ones.
		In any case, fragmenting for security increases latency inside the system.
		
		Nevertheless, a good fragmentation technique must be combined with the parallelization of processing to reach an acceptable overall performance. It should also take into consideration the fact that, for a collection of fragments, uneven levels of confidentiality may justify different levels of protection.
		
		This last idea has been successfully developed by Qiu \cite{bib:han} for the selective encryption of images using a general purpose GPU.
		
		Last but not least, the fragmentation of structured data strongly depends on user guidance for the definition of the confidentiality levels and depends a lot on the nature of the dataset. Designing an algorithm for automatically or semi-automatically separating confidential data from non-sensitive pieces would make the storage process much faster and easier to use.
		
		\subsection{Future work}
		
		\begin{figure}[!t]
		\centering
		\includegraphics[width=0.7\textwidth]{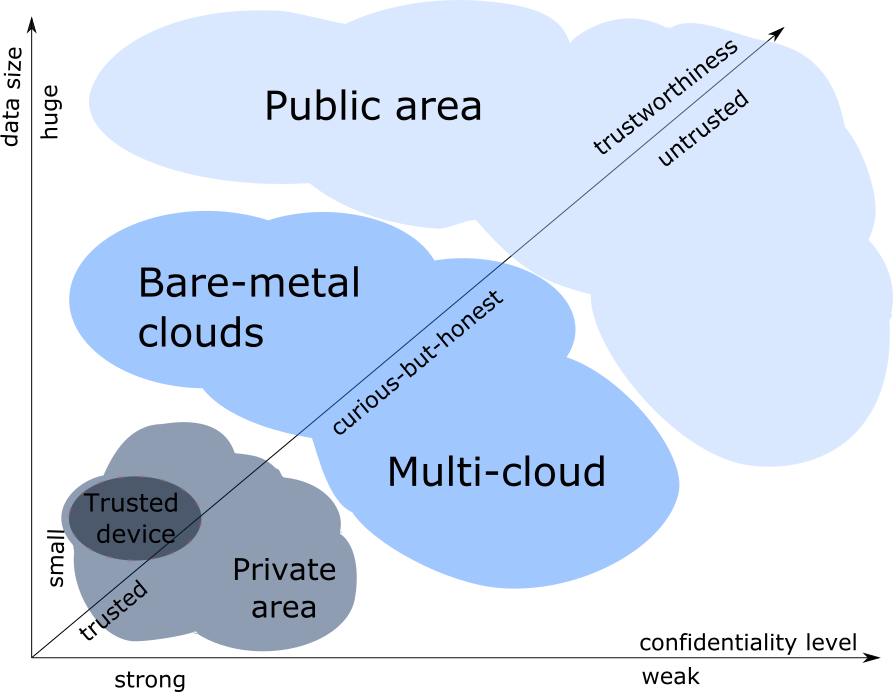}
		\caption{Data storage principle regarding data confidentiality level}
		\label{fig:cone}
		\end{figure}
		
		We plan to design a bitwise general scalable architecture that combines security with resilience while optimizing storage space and costs. This solution will leverage a large number of servers organized in public clouds or private distributed systems. At the same time, it will be cost effective, efficient in terms of performance and effective in terms of protection.
		
		Figure~\ref{fig:cone} illustrates one of the principles driving the architecture of our planned system. In the first approach, we divide the data into three categories based on the level of confidence: ultra-confidential, confidential and non-confidential. This multi-level
		security division will not focus on the problem of access control usual to such data categorization. It will be based on the risk that an owner is willing to take by exposing fragments of data in a public environment and their acceptance to keep the non-confidential data in an untrustworthy public storage space. The confidential data will be protected by placing it on secure devices.
		
		We distinguish three different storage spaces, adjusted to the chosen data security levels. Trusted area, most expensive and usually belonging to data owner or end user, will store the most sensitive data. The rest of the data will be dispersed to two types of clouds. The non-confidential part of information will be kept inside popular public clouds that use virtualization. Information that may potentially reveal a secret will go to bare-metals clouds, which provides more control of data location than the virtualization-based approach. Obviously, we are looking at minimizing the amount of data to be stored inside the most expensive zone and minimizing overall storage cost.
		
		At the end of our work process, we will concentrate on overcoming performance issues by a widespread use of parallel processing. As the final and most important step, we see data dispersion along levels of confidentiality.

		\bibliographystyle{abbrv}
		\bibliography{surveybib}

		\begin{landscape}
		
		\begin{table*}[ht]
	
		
		
		\caption{Comparison of most relevant bitwise fragmentation techniques. K - size of the encryption key, M - size of the dispersal matrix, H - size of the integrity canary, R - size of the added data inside a ramp scheme.}
		\centering
		\label{table:techniques}

		\begin{tabular}{ 
		>{\raggedright\arraybackslash}p{0.1\linewidth} 
		>{\raggedright\arraybackslash}p{0.2\linewidth}  
		>{\raggedright\arraybackslash}p{0.33\linewidth} 
		>{\centering\arraybackslash}p{0.15\linewidth}    
		>{\centering\arraybackslash}p{0.1\linewidth}    
		>{\centering\arraybackslash}p{0.1\linewidth}    
		}
	
		\hline \\  [-1ex]
		
		\textbf{Scheme} & \textbf{Security}  & \textbf{Performance} &  \textbf{Storage} & \textbf{Resilience}  & \textbf{Deduplication}    \\
		 
		\hline \\ [-1ex]
		
	SSS &  Information-theoretic & Cost of evaluating a value of polynomial of degree $k-1$ at $n$ points for each data chunk

	&  $nd$  & Yes & No  \\ \\ [-1ex]
		 
	XOR-split  & Information-theoretic & Cost of $k$ exclusive-or operations
	 & $kd$  & No & No   \\ \\ [-1ex]
		\\ [-1ex]    
		    SSMS      & Computational &
	Cost of: \begin{itemize}[leftmargin=*]
	\item  encryption
	\item  information disperal
	\item  perfect sharing scheme for key 
	\vspace{-1em} 
	\end{itemize} 	      
		      &  $\frac{nd}{k}+Kn$ & Yes & No   \\ \\ [-1ex]
		     CAONT-RS &  Computational &
	Cost of: \begin{itemize}[leftmargin=*]
	\item  producing the encryption key from hash
	\item  OAEP encryption
	\item  hashing data
	\item  systematic Reed-Solomon
	\vspace{-1em} 
	\end{itemize}
		     
		      & $\frac{n(d+H)}{k}+K$ & Yes & Yes  \\ \\ [-1ex]
		     AONT-RS & Computational & 
		     Cost of: \begin{itemize}[leftmargin=*]
	\item  AES encryption
	\item  hashing data
	\item  systematic Reed-Solomon
	\vspace{-1em} 
	\end{itemize} & $\frac{n(d+H)}{k}+K$ & Yes & No   \\ \\ [-1ex]
		\\ 
		 
		  
		   RSSS   & Perfect, then incremental &
	Cost of applying an IDA
	&  $\frac{d+R}{k}n$  & Yes  & Yes  \\ \\ [-1ex]
	
	CRSSS   & Computational, then incremental & 	
		\vspace{-1em} Cost of: \begin{itemize}[leftmargin=*]
	\item  producing fragments of R size using hashes
	\item  applying an IDA
	\vspace{-1em} 
	\end{itemize} 
	&   $\frac{d+R}{k}n$  & Yes  & Yes  \\ \\ [-1ex]	
		   IDA   & Incremental &
		\vspace{-1em}  Cost of multiplying data by a matrix $k\times n$
	
	&   $\frac{nd}{k}$  & Yes  & Yes \\ [45pt] \hline
		\end{tabular}
		
		
		
		\end{table*}
		
		\end{landscape}
		
				\begin{landscape}
		\begin{table*}[ht]
		
		
		
		\caption{Comparison of brute-force data dispersion academic storage systems. Blocks for compromise - number of fragments needed for data recovery,  \textit{b} -number of Byzantine faulty servers, \textit{d} - initial data size, \textit{k} - minimum number of fragments required for data reconstruction,  \textit{l} -number of leakage faulty servers, \textit{n} - total number of fragments, \textit{n\tiny{1}} -number of fragments in the first split, \textit{n\tiny{2}} -number of fragments in the second split, \textit{N} - number of servers, \textit{p} - minimum number of fragments that reveals any information about data, \textit{r} -numbe r of servers in a row in the logical grid.}
		\centering
		
		\label{table:tabelka1}
		\vspace{-1em} 
		\begin{tabular}{
		>{\raggedright\arraybackslash}p{0.12\linewidth}
		>{\raggedright\arraybackslash}p{0.12\linewidth}>{\raggedright\arraybackslash}p{0.15\linewidth}
		>{\raggedright\arraybackslash}p{0.15\linewidth}>{\raggedright\arraybackslash}p{0.15\linewidth}
		>{\raggedright\arraybackslash}p{0.15\linewidth}>{\raggedright\arraybackslash}p{0.15\linewidth}}
		\hline \\  [-2ex]
		& \textbf{Delta-4} & \textbf{PASIS}  & \textbf{POTSHARDS} & \textbf{GridSharing} & \textbf{DepSky} & \textbf{CDStore} \\
		\hline \\  [-2ex]
		
		Data  secrecy \newline
	& \textbf{Delta-4}	 &  Various threshold schemes adapted to information type & Perfect sharing scheme: XOR splitting $n_1$ of $n_1$ &  Perfect sharing scheme: XOR splitting & Krawczyk's SSMS  (with AES) &  CAONT-RS \\ \\ [-1ex]
		
		Data resilience, availability \newline
		& \textbf{Delta-4} & Threshold schemes \textit{p-k-n} & Threshold splitting $k$ of $n_2$  &  Fragments replication & Krawczyk's SSMS & Reed-Solomon error correction \\ \\ [-1ex]
		 
		Key management \newline 
		& \textbf{Delta-4}  &  Depending on the chosen protection method &  Keyless  &  Keyless  & Keyless & Keyless  \\ \\ [-1ex]
		 
		Integrity and recovery \newline 
		 & \textbf{Delta-4} & Repair agents, auditing & Algebraic signatures  & Voting system & Signatures (RSA) & Fingerprints (SHA-256) \\ \\ [-1ex]
		 
		Defragmentation  \newline 
		& \textbf{Delta-4}  &  Directory service maps objects and fragments. Fragment name contains the node location and the local name. 
		 
		 &  User knows shards indexes and data decomposition. In case of indexes loss: use of approximate pointers.  
		 
		 &  Voting system: user asks multiple servers for one fragment and chooses the most appropriate fragment from received answers. 
		 
		 &  User asks $k$ clouds for fragments with most recent data version.  
		 
		 &  CDStore client contacts CDStore servers that store information about dispersed data fragments. 
		 
		 \\ \\ [-1ex]
		 
		Trusted element \newline 
		& \textbf{Delta-4} & PASIS agent integrated within the client system & Data transformation component &  Client  & Client & CDStore client \\ \\ [-1ex]
		 
		Blocks for compromise & \textbf{Delta-4} & $k$ of $n$ & $k$ of $n_2$ for each of  $k$ of $n_1$  fragments & $n$  &  $k$ of $n$ &  $k$ of $n$ \\  \\ [-1ex]

		Storage space & \textbf{Delta-4} & $nd$ & $n_1n_2d$   & $ {{r}\choose{l+b}}(r-l-b)\frac{Nd}{r} $   & $\frac{n}{k}d$ & $\frac{n}{k}d$, deduplication \\ [4pt] \hline
		
		\end{tabular}
		
		
		
		\end{table*}
		\end{landscape}

		\begin{table*}[ht]
		
		\caption{Comparison of brute-force data dispersion commercial storage systems. Blocks for compromise - number of fragments needed for data recovery,  \textit{d} - initial data size, \textit{k} - minimum number of fragments required for data reconstruction, \textit{n} - total number of fragments.}
		\centering
		
		\label{table:tabelka2}
		
		\begin{tabular}{
		>{\raggedright\arraybackslash}p{0.2\linewidth}
		>{\raggedright\arraybackslash}p{0.2\linewidth}
		>{\raggedright\arraybackslash}p{0.2\linewidth}
		>{\raggedright\arraybackslash}p{0.2\linewidth}
		p{0.2\linewidth}}
		
		\hline \\ [-1ex]
		 & \textbf{IBM Cloud Object Storage} & \textbf{Symform} &\textbf{SecureParser}{\small{\small\textregistered}}\ \\
		\hline \\ [-1ex]
		
		Data encryption,  secrecy \newline 
		 &  AONT-RS with AES-256 &  AES-256  &  Data shredding, AES-256 \\  \\ [-1ex]
		
		Key management \newline 
		 &   Key stored within the data &   Key stored in the trusted element & Key stored within the data \\ \\ [-1ex]
		 
		Data resilience, availability \newline 
		 & Reed-Solomon error correction \newline \textit{k} of \textit{n} &  Reed-Solomon error correction \newline 64 of 96 & Threshold splitting \textit{k} of \textit{n} \\ \\ [-1ex]
		
		Integrity and recovery \newline 
		 & SHA-256 & SHA-256 & Authentication value inside each fragment \\ \\ [-1ex]
		
		Defragmentation  \newline 
		 &  Maps in Accesser switches indicate the location of fragments and the optimal storage node to contact. &  Map stored in the Symform Cloud Control.  & Map stored inside the Parsed File System. \\ \\ [-1ex]
		
		Trusted element \newline  & Accessers integrated or not within the client appliance & Symform Cloud Control central component &  Integrated within different elements of the client architecture \\ \\ [-1ex]
		
		Blocks for compromise & \textit{k} out of \textit{n} & 64 out of 96 & \textit{k} out of \textit{n} including all mandatory shares \\  \\ [-1ex]
		Storage space  & $ \frac{n}{k}d $ & $ (1.5)d $  & $ \frac{n}{k}d $ \\ [4pt] \hline
		\end{tabular}
		
		\end{table*}

		\end{document}